\documentstyle[twocolumn,prl,floats,aps]{revtex}
\begin{document}
\twocolumn[\hsize\textwidth\columnwidth\hsize\csname @twocolumnfalse\endcsname
\title{Non-exclusion statistics: a generalization of Bose-Einstein's principle}
\author{Yupeng Wang}
\address{Institute of Physics \& Center for Condensed Matter Physics, Chinese Academy of Sciences, 
Beijing 100080, People's Republic of China}

\maketitle
\date{\today}
\maketitle
\begin{abstract}
By constructing the super-particle representation of the free boson gas, we propose a new statistics in
which the particles are non-exclusive. This statistics can be considered as a generalization
of Bose-Einstein's. The possible condensation of this statistical system is studied. It is found
that the chemical potential below the condensation temperature is linearly proportional to
the temperature rather than a constant. With an proper choice
of the exclusion factors $\gamma_l$, Hadane-Wu's fractional statistics is retrieved in this representation.
\end{abstract}
\pacs{05.30.-d, 05.70.-a, 05.30.jp}]
\narrowtext
Bose-Einstein statistics and Fermi-Dirac statistics are two paradigms of quantum statistical mechanics.
The former is non-exclusive, i.e., an arbitrary number of particles (bosons) are allowed to occupy a
single-particle quantum state, while the latter is exclusive and in maximum only one particle (fermion) is allowed to 
occupy a single-particle quantum state. In a condensed matter, however, the interaction may change the statistics of
the quasi-particles drastically. A well known example is the one-dimensional hard-core boson lattice, which
is identical to a free fermion lattice via the Jordan-Wigner transformation. That means the infinite on-site
repulsion  changes the statistics of the bosons to a Fermi-Dirac one. In fact, any small repulsive
interaction in a one-dimensional boson system may drive it to behave as a Fermi system (in the sense that the
system has a Fermi surface), as shown by Yang and Yang in an integrable model.\cite{1} Motivated by the study
of spinons in one-dimensional antiferromagnets,\cite{2,3,4} Haldane formulated a general exclusion statistics\cite{5}
without specific reference to spatial dimension. With Haldane's definition, the maximum occupation number
of particles in a single-particle quantum state is just the inversion of the statistical factor $g$, as shown by Wu.\cite{6}
Therefore this definition can be considered as a generalization of Pauli's exclusion principle. When applied
to fractional quantum Hall effect, it coincides with the standard $2D$ anyon definition\cite{7} in terms of the
braiding of particle trajectories, though in general they are different. There are indeed some systems in
which the interactions among the bare particles can be completely absorbed into the quasi-particles with which
the systems behave as an ideal statistical ones. Concrete examples are the Calogero-Sutherland model\cite{8,9}
and the Haldane-Shastry spin chain.\cite{3,4} 
\par
Up to date, a successful Bose liquid theory corresponding to Landau's Fermi liquid theory has not been 
established. The main difficulty lies in the absence of an energy cutoff (counterpart of the Fermi energy in
a fermion system). Though an arbitrarily small repulsive interaction may drive a one-dimensional boson system
to behave as a non-ideal exclusion system\cite{10,11} or a Luttinger liquid,\cite{12} the superfluidity of the
liquid $^4$He and the condensation of some alklai atoms strongly indicate that the interactions of bosons
may not induce an extremely exclusion in three dimensions. In this letter, we formulate an ideal non-exclusion
statistics beyond that of Bose-Einstein. Comparing to Haldane's exclusion statistics, this new statistics
can be considered as a generalization of Bose-Einstein's principle. The deviation from Bose-Einstein 
statistics can be due to some interaction which is supposed to be absorbed into the quasi-particles completely.
\par  
For a free boson system, an arbitrary number of particles occupying a single-particle quantum state is allowed.
In addition, the physical quantities of these particles are additive, i.e., if $n$ particles are in the quantum
state labeled by the momentum $\vec k$ with single-particle energy $\epsilon(\vec k)$, their total momentum
and energy take the value of $n\vec k$ and $n\epsilon(\vec k)$, respectively. Therefore, $l$ bosons with mass $M$ in a
single-particle state $\vec k$ can be treated as a super-particle of mass $lM$, momentum $l\vec k$ and energy
$l\epsilon(\vec k)$. We call it simply a rank-$l$ super-particle. Obviously, a rank-$l$ super-particle can
also be treated as the combination of some rank-$l_\alpha$ super-particles with $\sum_\alpha l_\alpha=l$. If
we treat all types of super-particles independently, the many-particle quantum states must be over counted.
Hence an intrinsic exclusion among the super-particles defined in the same single-particle quantum
state should be taken into account to give a correct statistics. In the thermodynamic limit, we define the
density of rank-$l$ super-particles in the single-particle phase space as $\rho_l(\vec k)$ and the corresponding hole density
as $\rho_l^h(\vec k)$, respectively. The density of particles (rank-$1$) is given by $n(\vec k)=\sum_{l=1}^\infty
l\rho_l(\vec k)$. The intrinsic exclusion of the super-particles is expressed formally as
\begin{eqnarray}
\rho_l^h(\vec k)=\rho_l^0-\sum_{m=1}^\infty f_{lm}\rho_m(\vec k),
\end{eqnarray}
where $\rho_l^0$ denotes the density of rank-$l$ holes in the absence of particles and $f_{lm}$ are the intrinsic exclusion
factors. Since the volume of a boson in the phase space does not depend on its mass and we measure the densities
in the phase space of rank-$1$ particles, we readily have $\rho_l^0=l\rho_1^0$ with $\rho_1^0=1/(2\pi\hbar)^d$
in $d$ spatial dimensions as usual. The linear dependence of $\rho_l^h$ and $\rho_m$ is up to now still an
assumption. We note that if Eq.(1) gives a correct description of a free boson system, the exclusion
factors $f_{lm}$ ought to be dimension independent. To show them explicitly, let us consider a one-dimensional
system consisting of $N$ interacting bosons with $\delta$-potential. The Hamiltonian reads
\begin{eqnarray}
H=-\sum_{j=1}^N\frac{\partial^2}{\partial x_j^2}-2c\sum_{i<j}\delta(x_i-x_j).
\end{eqnarray}
This model was shown to be exactly solvable\cite{10} for arbitrary real $c$ and for $c\to0$, it is reduced to a 
free boson system. The eigenstates of Eq.(2) are described by $N$ momenta $\{k_j\}$ which are determined by
the so-called Bethe ansatz equations\cite{10}
\begin{eqnarray}
e^{ik_jL}=-\prod_{m=1}^N\frac{k_j-k_m-ic}{k_j-k_m+ic},
\end{eqnarray}
with the eigenenergy of Eq.(2) as $E=k_j^2$, where $L$ is the length of the system. For $c\to0^-$, it has
been demonstrated\cite{1} that the Bethe solution gives exactly the Bose-Einstein statistics. For $c>0$, Eq.(3)
allows string solutions with
\begin{eqnarray}
k_{j,\alpha}^l=k_{j}^l+\frac i2(l+1-2\alpha)c, {~~~~~~~}\alpha=1,2,\cdots,l.
\end{eqnarray}
As $c\to 0^+$, the strings are squeezed to the real axis and hence can be treated as the super-particles. Substituting
the squeezed-string solutions into Eq.(3) and take the logarithm, we readily obtain
\begin{eqnarray}
lk_j^lL=2\pi I_j^l
\nonumber\\
+\pi\sum_{i=1}^N\sum_{m=1}^\infty[2min\{l,m\}-\delta_{l,m}]\theta(k_j^l-k_i^m),
\end{eqnarray}
where $\theta(x)$ is the sign function; $I_j^l$ are integers (if $N$ odd) or half odd integers (if $N$ even).
Notice above we have taken the Planck constant $\hbar$ and the double mass $2M$ as our unit. Without the prefactor
$l$, Eq.(5) is just the equation of the Haldane-Sharstry spin chain obtained by Ha and Haldane.\cite{13}
In the continuum limit $L\to\infty$, put $I_j^l\to I_l(k)$. Then $\rho_l(k)+\rho_l^h(k)=dI_l(k)/dk$ as pointed
out by
Yang and Yang.\cite{1} In this case, we derive out  Eq.(1) from Eq.(5) in one dimension with
\begin{eqnarray}
f_{lm}=2min\{l,m\}.
\end{eqnarray}
To verify further the validity of Eq.(1) with Eq.(6) for an ideal Bose gas, we study the thermodynamics
of such a system. The density of thermal potential of this system can be expressed formally as
\begin{eqnarray}
f=\sum_{l=1}^\infty\int l[\epsilon(\vec k)-\mu]\rho_l(\vec k)d{\vec k}-T\int s(\vec k)d\vec{k},
\end{eqnarray}
where $\mu$ denotes the chemical potential, $\epsilon(\vec k)$ is the energy of a single particle 
and $s(\vec k)$ is the entropy density which reads\cite{1}
\begin{eqnarray}
 s(\vec k)=\sum_{l=1}^\infty[(\rho_l(\vec k)+\rho_l^h(\vec k))\ln(\rho_l(\vec k)+\rho_l^h(\vec k))
 \nonumber\\
-\rho_l(\vec k)\ln\rho_l(\vec k)-\rho_l^h(\vec k)\ln\rho_l^h(\vec k)].
\end{eqnarray}
By minimizing the thermal potential via $\delta\rho_l(\vec k)$ with Eq.(1) and Eq.(6), we obtain
\begin{eqnarray}
f=-\frac T{(2\pi)^d}\sum_{l=1}^\infty\int l\ln[1+\eta_l^{-1}(\vec k)]d{\vec k},
\end{eqnarray}
where $d$ is the dimension of the system and $\eta_l(\vec k)\equiv \rho_l^h(\vec k)/\rho_l(\vec k)$ and at thermal equilibrium satisfy the relations
\begin{eqnarray}
\ln[1+\eta_l(\vec{k})]=\frac{l[\epsilon(\vec k)-\mu]} T+\sum_{m=1}^\infty f_{lm}\ln[1+\eta_m^{-1}(\vec k)].
\end{eqnarray}
After some manipulations, Eq.(10) can be converted to the following algebraic equations
\begin{eqnarray}
\eta_l^2(\vec k)=(1+\eta_{l+1}(\vec k))(1+\eta_{l-1}(\vec k)),
\end{eqnarray}
with the boundary 
conditions  $\lim_{l\to\infty}\eta_l(\vec k)/l=[\epsilon(\vec k)-\mu]/T$ and $\eta_0\equiv 0$. The
solution of Eq.(11) is\cite{14}
\begin{eqnarray}
\eta_l(\vec k)=\frac{\sinh^2[(l+1)\frac{\epsilon(\vec k)-\mu}{2T}]}{\sinh^2\frac{\epsilon(\vec k)-\mu}{2T}}-1.
\end{eqnarray}
From Eq.(10) we have
\begin{eqnarray}
\sum_{l=1}^\infty l\ln(1+\eta_l^{-1}(\vec k))\nonumber\\
=\frac12\lim_{N\to\infty}[\ln(1+\eta_N(\vec k))-N\frac{\epsilon(\vec k)-\mu}T]\\
=-\ln(1-e^{-\frac{\epsilon(\vec k)-\mu}T}).\nonumber
\end{eqnarray}
This allows us to rewrite the density of the thermal potential as
\begin{eqnarray}
f=\frac T{(2\pi)^d}\int \ln(1-e^{-\frac{\epsilon(\vec k)-\mu}T})d{\vec k}.
\end{eqnarray}
It is just the expected one for a free boson gas in $d$ dimensions. Therefore, the super-particle description
Eq.(1) with Eq.(6) is indeed a correct one for the ideal boson gas in the thermodynamic limit.
\par
For an interacting Bose system, the total energy will be lifted by the repulsions among the particles. That
means the repulsive interaction in some sense may induce an exclusion effect. Suppose we introduce the
interaction in the following way: (i)There is a one-to-one correspondence between the quasi-super-particles
and the super-particles of the free boson system; (ii)The interactions are completely absorbed into the
quasi-super-particles and induce further exclusion effect. This hypothesis defines the quasi-super-particle
distribution as
\begin{eqnarray}
\rho_l^h(\vec k)=\frac l{(2\pi)^d}-\sum_{m=1}^\infty f_{lm}\rho_m(\vec k)-\gamma_ln(\vec k).
\end{eqnarray}
Note in the above definition, we choose the exclusion term proportional to the occupation density $n(\vec k)$.
$\gamma_l$ are some positive constants which measure the strength of the exclusion effect. A simple choice
of the statistical factors is to take all of them to be an unique constant, i.e., $\gamma_l=\gamma$.
This definition is very similar to Haldane's but in the representation of quasi-super-particles. This however
allows us to construct an ideal statistics without complete exclusion. The density of the thermodynamic potential in our
case is still expressed as
\begin{eqnarray}
f=\sum_{l=1}^\infty\int l[\epsilon(\vec k)-\mu]\rho_l(\vec k)d{\vec k}\nonumber\\
-T\sum_{l=1}^\infty\int\{[\rho_l(\vec k)+\rho_l^h(\vec k)]\ln[\rho_l(\vec k)+\rho_l^h(\vec k)]\\
-\rho_l(\vec k)\ln\rho_l(\vec k)-\rho_l^h(\vec k)\ln\rho_l^h(\vec k)\}d{\vec k}.\nonumber
\end{eqnarray}
Notice that $\delta\rho_l^h(\vec k)$ and $\delta\rho_m(\vec k)$ are not independent. From Eq.(15) we have
\begin{eqnarray}
\delta\rho_l^h(\vec k)=-\sum_{m=1}^\infty(f_{lm}+m\gamma)\delta\rho_m(\vec k).
\end{eqnarray}
By minimizing the density of the thermodynamic potential $f$ with respect to $\rho_l(\vec k)$, i.e.,
 $\delta f/\delta\rho_l(\vec k)=0$, we obtain that
 \begin{eqnarray}
 \ln[1+\eta_l(\vec k)]=\frac{l[\epsilon(\vec k)-\mu]}{T}\nonumber\\
 +\sum_{m=1}^\infty(f_{lm}+l\gamma)\ln[1+\eta_m^{-1}(\vec k)].
 \end{eqnarray}
With the identities
\begin{eqnarray}
\frac 12(f_{l+1m}+f_{l-1m})=f_{lm}-\delta_{lm},{~~~~}l>1\nonumber\\
\frac 12f_{2m}=f_{1m}-\delta_{1m},
\end{eqnarray}
Eq.(18) can still be transformed into Eq.(11) but with different boundary conditions
\begin{eqnarray}
\eta_0(\vec k)\equiv 0,\nonumber\\
\lim_{l\to\infty}\frac{\eta(\vec k)}{l}=\frac{\epsilon(\vec k)-\mu}T+\gamma \ln Z(\vec k),
\end{eqnarray}
where
\begin{eqnarray}
\ln Z(\vec k)\equiv\sum_{m=1}^\infty \ln[1+\eta_m^{-1}(\vec k)].
\end{eqnarray}
The solutions of $\eta_l(\vec k)$ in the present case read
\begin{eqnarray}
\eta_l(\vec k)=\frac{\sinh^2(l+1)[\frac{\epsilon(\vec k)-\mu}{2T}+\frac \gamma 2\ln Z(\vec k)]}
{\sinh^2[\frac{\epsilon(\vec k)-\mu}{2T}+\frac \gamma 2\ln Z(\vec k)]}-1.
\end{eqnarray} 
From the $l=1$ case of Eq.(18) we derive the following relation for the function $Z(\vec k)$
\begin{eqnarray}
[Z(\vec k)-1]Z^\gamma(\vec k)=e^{\frac{\mu-\epsilon(\vec k)}T}.
\end{eqnarray}
From the $l\to\infty$ case of Eq.(18) we have the relation
\begin{eqnarray}
\sum_{l=1}^\infty l\ln[1+\eta_l^{-1}(\vec k)]=\nonumber\\
=\frac12\lim_{N\to\infty}[\ln(1+\eta_N(\vec k))-N\frac{\epsilon(\vec k)-\mu}T-N\gamma\ln Z(\vec k)]\\
=-\ln[2-Z(\vec k)],\nonumber
\end{eqnarray}
and therefore the density of the thermodynamic potential can be rewritten as
\begin{eqnarray}
f=\frac 1{(2\pi)^d}\int\ln[2-Z(\vec k)]d{\vec k}.
\end{eqnarray}
Since all $\eta_l(\vec k)$ are non-negative, from the definition we know that $Z(\vec k)\geq 1$. In addition,
$Z(\vec k)\leq 2$ to keep $f$ to be real. From Eq.(23), it is easily to derive
\begin{eqnarray}
T\frac{\partial Z(\vec k)}{\partial\mu}=\frac{Z(\vec k)-1}{1+\gamma(1-Z^{-1}(\vec k))}.
\end{eqnarray}
With the relation $\int n(\vec k)d{\vec k}=-T\partial f/\partial\mu$ we get the single-particle distribution
function $n(\vec k)$ as
\begin{eqnarray}
n(\vec k)=\frac 1{(2\pi)^d}\frac {Z(\vec k)-1}{[2-Z(\vec k)][1+\gamma(1-Z^{-1}(\vec k))]}.
\end{eqnarray}
As $\gamma=0$, $n(\vec k)$ is just the Bose-Einstein's distribution function. For $Z(\vec k)\to 2$, $n(\vec k)\to
\infty$. Therefore, Eq.(15) can be considered as a generalization of Bose-Einstein's statistics. It is interesting
to study the condensation behavior of the present system. As in the ideal boson gas, condensation occurs on
the ${\vec k}=0$ state, if there is any. Note that $Z(\vec k)$ is a decreasing function of $\mu$. When the temperature
is lowered, $\mu$ tends to a critical value with which $Z(0)=2$, corresponding to the singular point of $n(\vec k)$.
Obviously, $\mu(T)=\gamma T\ln2$ below the critical temperature $T_c$. For a given density of particles $n$ and
$\epsilon(\vec k)=k^2$, $T_c$ is determined by
\begin{eqnarray}
T_c^{-\frac d2}=\frac1{4\pi n}\int_0^\infty\frac{[z(x)-1]x^{\frac d2-1}dx}{[2-z(x)][1+\gamma(1-z^{-1}(x))]},
\end{eqnarray}
with
\begin{eqnarray}
[z(x)-1]z^\gamma(x)=2^\gamma e^{-x}.
\end{eqnarray}
For $x\to0$, $2-z(x)\sim x$, the integral in Eq.(28) is divergent for $d\leq2$, which indicates the absence of 
condensation at finite temperature. However, the finite-temperature condensation indeed occurs for $d>2$. A
simple analysis leads to $\partial z(x)/\partial \gamma>0$ and $\partial {T_c}/\partial\gamma<0$, which indicates
that $T_c$ is decreasing with the increase of $\gamma$ and $T_c\to0$ for $\gamma\to\infty$.

An interesting issue is that a proper choice of $\gamma_m$ may even induce some exclusion statistics. For
example, we put $\gamma_m=mg$ with $g$ a positive constant. In this case, we have
\begin{eqnarray}
\delta\rho_l^h(\vec k)=-\sum_{m=1}^\infty(f_{lm}+lmg)\delta\rho_m(\vec k).
\end{eqnarray}
With the above relation to minimize the thermodynamic potential Eq.(16), we get
\begin{eqnarray}
 \ln[1+\eta_l(\vec k)]=\frac{l[\epsilon(\vec k)-\mu]}{T}\nonumber\\
 +\sum_{m=1}^\infty(f_{lm}+lmg)\ln[1+\eta_m^{-1}(\vec k)].
 \end{eqnarray}
The solutions of $\eta_l(\vec k)$ take the following form
\begin{eqnarray}
\eta_l(\vec k)=\frac{\sinh^2(l+1)[\frac{\epsilon(\vec k)-\mu}{2T}+\frac \gamma 2\ln\zeta(\vec k)]}
{\sinh^2[\frac{\epsilon(\vec k)-\mu}{2T}+\frac \gamma 2\ln\zeta(\vec k)]}-1,
\end{eqnarray}
where
\begin{eqnarray}
\ln\zeta(\vec k)=\sum_{m_1}^\infty m\ln[1+\eta_m^{-1}(\vec k)],
\end{eqnarray}
and the thermodynamical potential in this case reads
\begin{eqnarray}
f=-\frac1{(2\pi)^d}\int\ln\zeta(\vec k)d{\vec k}.
\end{eqnarray}
With the relation
\begin{eqnarray}
\ln\zeta(\vec k)=
\frac12\lim_{N\to\infty}\{\ln[1+\eta_N(\vec k)]\nonumber\\
-N\frac{\epsilon(\vec k)-\mu}T-Ng \ln\zeta(\vec k)\},
\end{eqnarray}
we get
\begin{eqnarray}
[1-\zeta^{-1}(\vec k)]\zeta^g(\vec k)=e^{\frac{\mu-\epsilon(\vec k)}T}.
\end{eqnarray}
Define
\begin{eqnarray}
Y(\vec k)\equiv\frac1{\zeta(\vec k)-1}.
\end{eqnarray}
Eq.(36) can be transformed into
\begin{eqnarray}
Y^g(\vec k)[1+Y(\vec k)]^{1-g}=e^{\frac{\epsilon(\vec k)-\mu}T},
\end{eqnarray}
and 
the distribution function $n(\vec k)$ reads
\begin{eqnarray}
n(\vec k)=\frac T{(2\pi)^d}\frac{\partial\ln \zeta(\vec k)}{\partial \mu}=\frac1{(2\pi)^d}\frac1{Y(\vec k)+g}.
\end{eqnarray}
Eq.(38) and Eq.(39) are just the results of Haldane's exclusion statistics derived by Wu.\cite{6}
\par
In conclusion, by constructing the super-particle representation of the free boson gas, we introduce
a non-exclusion statistics of many-particle systems. the Bose-Einstein condensation of this system is
also studied. Unlike in the free boson gas, it is found that the chemical potential below the transition
temperature is linearly proportional to the temperature. Haldane-Wu's fractional statistics is perfectly
retrieved in this representation. This strongly suggests that with the super-particle representation,
the non-exclusion statistics and the exclusion statistics can be constructed in an uniform way.
\par
This work was supported by NSFC under grant No. 19825112.

\end{document}